\documentclass[letterpaper,12pt]{article}
\usepackage{amssymb,latexsym,amsmath}

\usepackage{fullpage}
\usepackage{nicefrac}
\usepackage{mathrsfs}
\usepackage{slashed}

\usepackage[pdftex]{graphicx}
\usepackage[table,dvipsnames]{xcolor}
\usepackage{multirow}
\usepackage{rotating}
\usepackage{bbold}

\makeatletter \@addtoreset{equation}{section}
\makeatother

\begin{document}

\begin{titlepage}
	\thispagestyle{empty}
	\begin{flushright}
		\hfill{DFPD-12/TH/3}
	\end{flushright}
	
	\vspace{35pt}
	
	\begin{center} 
{ \LARGE{\bf Quantum corrections to \\ \vspace{10pt} broken $N=8$ supergravity}}		
		\vspace{50pt}
		
		{Gianguido~Dall'Agata and Fabio~Zwirner}
		
		\vspace{25pt}
		
		{{\it  Dipartimento di Fisica ed Astronomia ``Galileo Galilei''\\
		Universit\`a di Padova, Via Marzolo 8, 35131 Padova, Italy}
		
		\vspace{15pt}

and

		\vspace{15pt}
		
	        {\it   INFN, Sezione di Padova \\
		Via Marzolo 8, 35131 Padova, Italy}
		
		}		
		
		\vspace{40pt}
		
		{ABSTRACT}
	\end{center}

We show that the one-loop effective potential of spontaneously broken $N=8$ supergravity is calculable and finite at {\em all} classical four-dimensional Minkowski vacua without tachyons in the spectrum. The reason is that the supertraces of the quadratic and quartic mass matrices vanish along the classically flat directions: ${\rm Str} \, {\cal M}^2 = {\rm Str} \, {\cal M}^4 =0$. We also show that ${\rm Str} \, {\cal M}^6 = 0$ but ${\rm Str} \, {\cal M}^8 > 0$ in a broad class of vacua with broken supersymmetry on a flat background, which includes all those explicitly identified so far. We find analytical and numerical evidence that the corresponding one-loop effective potential is negative definite.
	
	\vspace{10pt}

\end{titlepage}

\baselineskip 6 mm



\section{Introduction and summary} 
\label{sec:introduction}

One of the most intriguing aspects of four-dimensional quantum field theories with global or local supersymmetry is their better and better ultraviolet behaviour as the number $N$ of supersymmetries increases. Theories with maximal global supersymmetry, $N=4$ super-Yang-Mills theories, are known to be perturbatively finite, indeed (super-)conformally invariant at the quantum level. The theory with maximal local supersymmetry, $N=8$ supergravity, has recently revealed an unexpectedly good behaviour at higher loops, when quantized in its ungauged version on a flat background, and its possible finiteness is under intense discussion  (for a recent review and references, see e.g. \cite{Bern:2011qn}). 

In contrast with spontaneously broken gauge theories, whose ultraviolet behaviour is the same as in the unbroken phase, spontaneously broken extended supergravities do not share in general the ultraviolet properties of their unbroken versions. The reason is that, to break supersymmetry spontaneously, we must deform the theory by gauging a subgroup of its global symmetries, and this operation breaks some of the global symmetries and introduces new terms in the supersymmetry transformations and in the Lagrangian. The goal of the present work is to match the impressive recent advances on the ultraviolet behaviour of ungauged $N=8$ supergravity with some new insight on the one-loop corrections to its gauged version, focusing on those gaugings that lead to classical Minkowski vacua with fully broken supersymmetry.

Powerful indicators of the ultraviolet behaviour of supersymmetric theories are the supertraces of the even powers of the field-dependent mass matrices, defined by 
\begin{eqnarray}
{\rm Str} \, {\cal M}^{2k} & \equiv & \sum_a (2J_a +1) \, (-1)^{2J_a} \,  (M^2_a)^k 
\nonumber \\ & = & 
{\rm Tr} \, [{\cal M}_{(0)}^2]^k - 2 \ 	{\rm Tr} \, [{\cal M}_{(1/2)}^2]^k + 3 \ {\rm Tr} \, [{\cal M}_{(1)}^2]^k 
- 4 \ {\rm Tr} \, [{\cal M}_{(3/2)}^2]^k \, , 
\label{Supertrace}
\end{eqnarray}
where $k=0,1,2,\ldots$, the index $a$ runs over the different particles in the spectrum, $M^2_a$  and $J_a$ are the corresponding squared-mass eigenvalues and spins. More explicitly, the second line of the above equation refers to an arbitrary theory with any number of massive spin-3/2 gravitinos, spin-1 gauge bosons, spin-1/2 Weyl fermions and spin-0 real scalars. Of course, massless particles do not contribute to the supertraces. It is easy to see \cite{Coleman:1973jx, Weinberg:1973ua} that ${\rm Str} \, {\cal M}^0$, ${\rm Str} \, {\cal M}^2$ and ${\rm Str} \, {\cal M}^4$ control the quartic, quadratic and logarithmic divergences of the one-loop effective potential, respectively.  

It was observed long ago \cite{Zumino:1974bg} that in any theory with exact supersymmetry all the supertraces vanish on any Minkowski background. In the case of broken supersymmetry, the issue is more subtle. The coefficient of the one-loop quartic divergence,  ${\rm Str} \, {\cal M}^0$, is always field-independent, and counts the number of bosonic minus fermionic degrees of freedom, thus it vanishes in all theories with linear representations of supersymmetry \cite{Zumino:1974bg}. The coefficient of the one-loop quadratic divergence, ${\rm Str} \, {\cal M}^2$, vanishes in renormalizable theories with $N=1$ global supersymmetry and no anomalous $U(1)$ factors in the gauge group \cite{Ferrara:1979wa}, and remains field-independent, but no longer vanishes, when explicit but soft supersymmetry-breaking terms are added \cite{Girardello:1981wz}. In $N=1$ theories with non-renormalizable interactions, such as $N=1$ supergravity, the general expression \cite{Grisaru:1982sr} for ${\rm Str} \, {\cal M}^2$  has a geometric interpretation and is in general non-vanishing, even if special superstring-related $N=1$ models \cite{Ferrara:1987jq, Ferrara:1994kg} are known where ${\rm Str} \, {\cal M}^2 = 0$ with broken supersymmetry. 

Little is known on the properties of the quadratic and higher supertraces in gauged extended ($N>1$) supergravities. A notable exception is what was, until recently, the only well-studied example of spontaneously broken $N=8$, $D=4$ supergravity on a classically flat background: the four-parameter gauging of Cremmer, Scherk and Schwarz (CSS) \cite{Cremmer:1979uq}. It was realized already in the original CSS paper, and explained in group-theoretical terms in \cite{Ferrara:1979fu}, that for the CSS gauging ${\rm Str} \, {\cal M}^2 = {\rm Str} \, {\cal M}^4 = {\rm Str} \, {\cal M}^6 =  0$, but ${\rm Str} \, {\cal M}^8 \ne 0$ if there is no residual supersymmetry on the vacuum.

Thanks to some recent developments \cite{DallAgata:2011aa} (see also \cite{invthe, Dibitetto:2011gm}), we can now explicitly explore the classical and quantum stability of a much broader class of $N=8$ gaugings leading to classical vacua with fully broken supersymmetry in Minkowski space. Moreover, the general formalism for gauged $N=8$ supergravity \cite{deWit:2002vt, de Wit:2007mt} allows us to generalize part of the results already known for the CSS gaugings to {\em arbitrary} gaugings of $N=8$ supergravity. Working in the general formalism of $N=8$ gauged supergravity, we will show that the supertraces of the quadratic and quartic mass matrices vanish at {\em all} classical four-dimensional Minkowski vacua with spontaneously broken $N=8$ supergravity: 
\begin{equation}
\label{result1}
{\rm Str} \, {\cal M}^2 = {\rm Str} \, {\cal M}^4 =0 \, .
\end{equation}
This makes the one-loop effective potential calculable and finite, for all tachyon-free constant field configurations along the classically flat directions,
\begin{equation}
\label{respot}
V_1 = \frac{1}{64 \, \pi^2} \, {\rm Str} \, \left( {\cal M}^4 \, \log {\cal M}^2 \right) \, .
\end{equation}
Notice that such one-loop effective potential has the same scaling dimension of a classical potential. We will then move to the study of all those gaugings for which the full mass spectrum depends only on the charges of the eight supersymmetry generators with respect to at least one unbroken U(1) factor in the gauge group. Notice that the class of gaugings so defined includes all the explicit gaugings considered so far that lead to classical vacua with fully broken supersymmetry on a flat background. We will show that for all these gaugings not only
\begin{equation}
{\rm Str} \, {\cal M}^6 = 0 \, , 
\label{result2}
\end{equation}
but also
\begin{equation}
{\rm Str} \, {\cal M}^8 > 0 \, . 
\label{result3}
\end{equation}
We will also observe that, in contrast with the positive semi-definite classical potential generated by the CSS gauging, the potentials associated with the new gaugings of \cite{DallAgata:2011aa}  can have tachyonic instabilities along their flat directions, and we will identify those flat directions that do not lead to tachyons in the classical spectrum. After specializing to this part of the classical moduli space, we will study the one-loop effective potential and find analytical and numerical evidence that for all these field configurations, as a consequence of Eqs.~(\ref{result1})--(\ref{result3}), it is negative definite: 
\begin{equation}
V_1 = \frac{1}{64 \, \pi^2} \, {\rm Str} \, \left( {\cal M}^4 \, \log {\cal M}^2 \right) < 0 \, . 
\label{result4}
\end{equation}
This means that at the one-loop level no locally stable vacua are known with fully broken supersymmetry and positive or vanishing vacuum energy. We stress that all the results mentioned above are highly non-trivial extensions of the known results valid for the CSS gauging: for general gaugings, and in particular for all the other explicit gaugings considered in this paper, we cannot rely on the group-theoretical explanation that was shown to be at the root of the supertrace formulas valid for the CSS gauging. 	

Our paper is organized as follows. After this introduction, in Section~2 we briefly recall the general formalism of gauged $N=8$ supergravity and the general form of the resulting mass matrices. The explicit form of the quadratic and linear identities, which guarantee the consistency of the gauging and are essential for proving the results of the following sections, are displayed in the Appendix. In Section~3 we describe the derivation of the general result of Eq.~(\ref{result1}). In Section~4 we spell out two assumptions that are fulfilled by all the presently known gaugings leading to classical Minkowski vacua with fully broken supersymmetry. Under these assumptions, we derive the results of Eqs.~(\ref{result2}), (\ref{result3}) and (\ref{result4}). We then discuss in some detail the explicit mass spectra of these models, as functions of the parameters defining the gaugings. We conclude in Section~5 with a list of open questions and an outlook.



\section{Gauged $N=8$ supergravity and its mass matrices} 
\label{sec:$N=8$ mass matrices}

\subsection{Gaugings} 
\label{sub:Gaugings}

$N=8$ supergravity in 4 dimensions is usually described in terms of a unique multiplet containing the graviton, 8 gravitinos, 28 vector fields, 56 spin 1/2 fields and 70 scalars. The scalar fields parametrize the coset manifold E$_{7(7)}/$SU(8), where E$_{7(7)}$ is the duality group leaving invariant the system of Bianchi identities and equations of motion of the ungauged theory \cite{deWit:1977fk, Cremmer:1978ds, Cremmer:1979up, de Wit:1982ig}, i.e.~of the theory with exact supersymmetry and vanishing potential constructed without introducing non-Abelian gauge interactions. Although generically the Lagrangian is not invariant under the action of such a duality group, different choices of duality frames give classically equivalent theories on-shell. This is no more true when the theory is gauged. By the gauging procedure up to 28 physical vector fields are coupled to charges assigned to the other fields according to their transformation properties under the global symmetries of the ungauged theory, and a subgroup $G$ of the E$_{7(7)}$ duality group is rendered local. As a result, the original theory is deformed by the addition of new couplings, non-trivial masses and a scalar potential. Although, at the Lagrangian level, supersymmetry is restored by modifying the supersymmetry transformation rules with the addition of new terms proportional to the gauge coupling constant, the appearance of a scalar potential makes possible the spontaneous breaking of supersymmetry and of the gauge group.

The group of local gauge transformations is defined by means of its generators $X_M$, which must form a subset of the generators of the algebra of duality symmetries $t_{\alpha} \in {\mathfrak e}_{7(7)}$. This choice is completely specified by the embedding tensor $\Theta$, which collects the full set of deformation parameters \cite{Nicolai:2000sc}: 
\begin{equation}
	X_M = \Theta_M{}^\alpha \, t_{\alpha}.
\end{equation} 
In $N=8$ supergravity $\Theta$ has an index $\alpha$ in the $\mathbf{133}$-dimensional adjoint representation of ${\mathfrak e}_{7(7)}$ and an index $M$ in its $\mathbf{56}$-dimensional fundamental representation. In fact, in four dimensions massless vector fields are dual to themselves, thus we could in principle use as gauge fields suitable  combinations of the 28 electric and 28 magnetic vectors in a given duality frame \cite{Gaillard:1981rj}. This fact, together with the requirements of 
supersymmetry and gauge invariance, imposes a series of constraints that the embedding tensor must satisfy to produce consistent gaugings. In general there are two sets of constraints: linear and quadratic. Linear constraints are implied by supersymmetry and in $N=8$ supergravity they restrict $\Theta$ to the $\mathbf{912}$ representation in the decomposition of the $\mathbf{56} \otimes \mathbf{133}$ product. Explicitly this means that
\begin{equation}
	X_{M\, [NP]} = 0 = X_{(MNP)} = X_{MN}{}^N = X_{NM}{}^N,
	\label{contractedquadraticconstraint}
\end{equation}
where $X_{MN}{}^P = (X_M)_N{}^P$ and the $M,N$ indices are raised and lowered with the Sp(56,$\mathbb R$) skew-symmetric invariant tensor $\Omega_{MN}$.
The quadratic constraints are equivalent to the invariance of the tensor $\Theta$ under the action of the generators of the local gauge symmetry and force the closure of the $G$ generators into an algebra\footnote{It should be noted that $X_{MN}{}^P$ may contain a symmetric part in the lower indices. In this case gauge invariance requires the introduction of tensor fields with suitable gauge transformations \cite{de Wit:2007mt}, so that the gauge structure is that of a free differential algebra rather than that of an ordinary Lie algebra \cite{Dall'Agata:2005mj,deWit:2005ub}. Also, the embedding tensor really forces to gauge a group whose \emph{adjoint} representation is embedded in the duality group. This is in general a quotient of the one generated by the representation of the vector fields with its maximal Abelian ideal \cite{Dall'Agata:2007sr}.}:
\begin{equation}
\,	[ X_M, X_N] = - X_{MN}{}^P X_P.
	\label{contractedquadraticconstraints}
\end{equation}
Once the linear constraints are taken into account, the quadratic constraints can also be expressed in an equivalent fashion by
\begin{equation}
	\Theta_M{}^\alpha \Theta_N{}^\beta \, \Omega^{MN} = 0,
\end{equation}
which makes clear that at most 28 vectors can be used as gauge fields and that there is always a symplectic frame where all the $\Theta^\alpha$ are electric.

The $\mathbf{912}$ possible deformations of $N=8$ gauged supergravities parametrized by the embedding tensor are in one-to-one correspondence with the elements of the shift matrices of the supersymmetry transformation rules of the fermions. These matrices are field-dependent objects that transform in definite representations of the $R$-symmetry group SU(8) and are determined by the decomposition of the $\mathbf{912}$ representation of E$_{7(7)}$ under SU(8). In detail, they can be constructed by contracting the gauge generators with the E$_{7(7)}/$SU(8) coset representatives ${\cal V}$ and by decomposing the resulting tensorial structure, called $T$-tensor, in its irreducible components. Since the coset representatives are matrices whose indices transform in the fundamental representation of E$_{7(7)}$ and in the $\mathbf{28} \oplus \overline{\mathbf{28}}$ of SU(8)  (${\cal V} = \{{\cal V}_M{}^{ij}, {\cal V}_{M\, ij}\}$), the $T$-tensor is defined by \cite{de Wit:1981eq}
\begin{equation}
T_i{}^{jkl} = {\cal V}^{-1\,kl\,M} {\cal V}^{-1}_{mi}{}^{N} X_{MN}{}^P {\cal V}_P{}^{mj}
= \frac{3}{2} \delta_i^{[k} A_1^{l]j} - \frac{3}{4}A_2{}_i{}^{jkl}
\label{Ttensor}
\end{equation} 
and by its complex conjugate. $A_1$ and $A_2$ describe its SU(8) irreducible components. $A_1$ is in the $\mathbf{36}$ representation of SU(8), so that $A_{1}^{ij} = A_1^{ji}$, while $A_2$ is in the $\mathbf{420}$ representation, so that $A_{2\,i}{}^{jkl} = A_{2\,i}{}^{[jkl]}$, $A_{2\,i}{}^{ijk}=0$. The complex conjugate of (\ref{Ttensor}) contains the representations $\overline{\mathbf{36}}$, described by $A_{1\,ij} = (A_1^{ij})^*$, and $\overline{\mathbf{420}}$, described by $A_{2\,i}{}^{jkl} = (A_2{}^i{}_{jkl})^*$. As already mentioned, these very same matrices define the modification of the spinor supersymmetry transformations that is relevant for the discussion of maximally symmetric vacua of the theory. With respect to the ungauged case, these terms shift the supersymmetry variations as follows:
\begin{eqnarray}
	\delta_g \psi_\mu^i &=& \delta_0 \psi_\mu^i + \sqrt{2} \, A^{ij} \, \gamma_\mu \epsilon_j\,,\\[2mm]
	\delta_g \chi^{ijk} &=& 	\delta_0 \chi^{ijk} - 2\, A_{l}{}^{ijk} \, \epsilon^l.
\end{eqnarray}

It is quite convenient to express also the scalar potential and the mass matrices of the gauged theory in terms of the same $A_1$ and $A_2$ tensors. We will discuss in detail the form of the mass matrices, which are at the base of our computations, in the next section.
Here we just give the form of the classical scalar potential
\begin{equation}
	V_0(\phi) =  \frac{1}{24} A_{n}{}^{jkl} {A{}^n}_{jkl} -\frac{3}{4} A_{kl} A{}^{kl},
	\label{Lambda}
\end{equation}
which determines the value of the cosmological constant of the classical vacua of the theory. Since we are interested in maximally symmetric solutions of the field equations with constant scalars, vanishing vector fields and vanishing fermions, imposing the vacuum condition is equivalent to finding critical points of the scalar potential. These are specified by solutions of the following selfdual system of $70$ complex equations \cite{deWit:1983gs}
\begin{equation}
	\label{vacuum}
	{A{}^m}_{[ijk} A_{l]m}
	+ \frac34 {A{}^m}_{n[ij} {A^n}_{kl]m}
	+ \frac{1}{24} \epsilon_{ijklabcd} \left({A{}_m}^{abc} A^{dm}
+ \frac34 {A{}_m}^{n ab} {A_n}^{cdm}\right) = 0.
\end{equation}


\subsection{Mass matrices} 
\label{sub:Mass matrices}

The computation of the supertrace mass formulas requires the knowledge of the spectrum of quadratic fluctuations at the maximally symmetric critical points described by Eq.~(\ref{vacuum}). The matrices determining such spectrum were computed in \cite{LeDiffon:2011wt} by linearizing the field equations and in what follows we will use their results and conventions (for more details on the conventions see also \cite{de Wit:2007mt}), although we will provide an alternative way of determining the masses of the spin 1/2 fields that will simplify dramatically our supertrace computations.

Before writing down the explicit formulas for the mass matrices, we recall that the fields of the $N=8$ supergravity multiplet fit in SU(8) representations that are best described by means of totally antisymmetric multi-indices $I = [i_1 \ldots i_n]$, where $i_1, \ldots i_n$ transform in the (anti-)fundamental representation of SU(8). The gravitinos are in the $\mathbf{8}$-dimensional fundamental representation: $\psi_\mu^i$. The 28 electric and 28 magnetic vector fields are in the $\mathbf{28} \oplus \overline{\mathbf{28}}$ representations: $\left\{A_\mu^{ij},A_{\mu\,ij}\right\}$. Spin 1/2 fields sit in the $\mathbf{56}$: $\chi^{ijk}$. Finally, scalars are described by complex combinations in the $\mathbf{70}$, $\phi^{ijkl}$, such that their real and imaginary parts satisfy opposite self-duality conditions $\phi_{ijkl} = \frac{1}{24} \, \epsilon_{ijklmnpq} \phi^{mnpq} = (\phi^{ijkl})^*$.

The mass matrices we are going to use in the following will be expressed using such  multi-indices and therefore, to avoid overcounting when taking traces, we should choose a proper normalization. Following \cite{LeDiffon:2011wt}, our mass matrices enter the equations of motion of a generic field $\varphi^I$ as
\begin{equation}
	\Box \varphi^I = M^{I}{}_J \varphi^J,
	\label{bosonicmass}
\end{equation}
for bosonic degrees of freedom and as
\begin{equation}
	\slashed \partial  \varphi^I =  M^{IJ} \varphi_J,
\end{equation}
for fermionic degrees of freedom. This means that the masses obtained by diagonalizing $M^I{}_J$, considering different entries only when $I \neq J$ after rearrangement of the multi-index, are rescaled by a factor $1/n!$. This however is the factor needed to obtain the correct value from the trace $M_I{}^I$, because we are summing each mass term $n!$ times. For instance, for $I = 12 = - 21$, $M^{ij}{}_{ij} = M^{12}{}_{12}+M^{21}{}_{21} = 2 M^{12}{}_{12}$. If $M^{12}{}_{12} = m^2/2$ we get the correct normalization for the equation of motion of the corresponding field $\Box \varphi^{12} = m^2 \varphi^{12}$. Taking note of this normalization issue, we can now provide the explicit expressions for the bosonic and fermionic mass matrices.

In $N=8$ supergravity there are 28 vector fields, but in a given symplectic frame, those used in the gauged theory may be a mixture of the 28 original electric fields $A^{ij}_\mu$ and of the 28 magnetic ones $A_{\mu\,ij}$. For this reason, the vector mass matrix is a square matrix with 56 rows and columns \cite{LeDiffon:2011wt}:
\begin{equation}
	{\cal M}_{(1)}^2 =
	\left(
	\begin{array}{cc}
	{\cal M}_{ij}{}^{kl} & {\cal M}_{ijkl}
	\\[2mm]
	 {\cal M}^{ijkl}&
	 {\cal M}^{ij}{}_{kl}
	\end{array}
	\right)
	\;,
	\label{M_vector}
\end{equation}
where
\begin{equation}
	{\cal M}_{ij}{}^{kl}
	= ({\cal M}^{ij}{}_{kl})^*= 
	-\frac16 A_{[i}{}^{npq} \delta_{j]}^{[k} A^{l]}{}_{npq}
	+\frac12 A_{[i}{}^{pq[k}A^{l]}{}_{j]pq}
\end{equation}
and
\begin{equation}
	{\cal M}_{ijkl} = ({\cal M}^{ijkl})^* =
	\frac1{36}  A_{[i}{}^{pqr} \epsilon_{j]pqrmns[k} A_{l]}{}^{mns}\,.
\end{equation}
Since 28 vector fields are not physical, when diagonalizing such matrix we should always find at least 28 null eigenvalues.

The scalar mass matrix is \cite{deWit:1983gs, LeDiffon:2011wt}
\begin{equation}
	\begin{array}{rcl}
	 {\cal M}_{ijkl}{}^{mnpq} &=&\displaystyle
	6\left(A_{[i}{}^{ab[m}A^{n}_{j|ab|}\delta_{kl]}^{pq]}\!-\!\frac14\,A_{t}{}^{s[mn}A^{|t|}_{s[ij}\delta_{kl]}^{pq]}\right)
	\\[3mm]
	&&{}\displaystyle
	+\left(\frac{5}{24}\,A_{a}{}^{bcd}A^{a}{}_{bcd}-\frac12A_{ab}A^{ab}\right) \delta^{mnpq}_{ijkl}
	\\[3mm]
	&&{}\displaystyle
	-\frac23\,A_{[i}{}^{[mnp} A^{q]}{}_{jkl]} 
	\;.
	\end{array}
	\label{scalarmassmatrix}
\end{equation}
An important point that should be taken into account when computing the masses of the scalar fields is that these fields satisfy the self-duality condition $\phi_{ijkl} = \frac{1}{24} \epsilon_{ijklmnpq} \phi^{mnpq} = (\phi^{ijkl})^*$. This means that different entries of the mass matrix (\ref{scalarmassmatrix}) contribute to the same mass term. In turn, this implies that we should be careful when computing traces. To separate the selfdual and anti-selfdual terms we can use the projectors $P_{\pm}{}_{ijkl}^{abcd} = \frac1{2}\left(\delta_{ijkl}^{abcd} \pm \frac1{24} \epsilon_{ijkl abcd}\right)$ (obviously we have to break SU(8) covariance).
We can then construct 
\begin{equation}
	{\cal M}_{(0)}^2 = \left(\begin{array}{cc}
	P_+ {\cal M} P_+ & 	P_+ {\cal M} P_- \\[2mm]
	P_- {\cal M} P_+ & 	P_- {\cal M} P_-
	\end{array}\right)
\end{equation}
and consider its traces.

The fermion mass matrices have somewhat simpler expressions. The gravitino mass matrix is simply proportional to the shift of the gravitinos in the supersymmetry transformation rule,  
\begin{equation}
	{\cal M}_{(3/2)}{}^{ij} = \sqrt{2}\,A^{ij} \, , 
\end{equation}
and the spin-1/2 mass matrix is proportional to the shift term in the corresponding supersymmetry transformation. However, since at a supersymmetry--breaking vacuum the super--Higgs mechanism will be at work, we need to evaluate the spin-1/2 mass matrix after projecting out the goldstinos that are eaten up by the massive gravitinos. This is obtained by diagonalizing the fermion mass matrices, redefining the gravitinos in a way that gets rid of the mixed term between them and the spin-1/2 fields. A quick inspection of the relevant part of the $N=8$ Lagrangian shows that the fermion mass terms are \cite{de Wit:2007mt}
\begin{equation}
	\label{Lmass}
	{\cal L}_{mass} = \frac{\sqrt2}{2} \,  A_{ij} \, \bar\psi_\mu^i \gamma^{\mu\nu} \psi_\nu^j + \frac16 \, A_i{}^{jkl} \, \bar \psi_\mu^i \gamma^\mu \chi_{jkl} + \frac{1}{2\cdot 6} M^{ijk,lmn} \bar \chi_{ijk} \chi_{lmn} + {\rm h.c.},
\end{equation}
where
\begin{equation}
	M^{ijk,lmn} = \frac{\sqrt2}{12} \epsilon^{ijkpqr[lm}\,A^{n]}{}_{pqr},
\end{equation}
so that the field equation for the spin-1/2 fields becomes
\begin{equation}
	\slashed\partial\, \chi^{ijk} = M^{ijk,abc} \chi_{abc} + \ldots \, .
\end{equation}
In the above equation, the dots contain further interaction terms, among which there is a term proportional to the spin 1/2 projection of the gravitinos $A_{m}{}^{ijk}\, \gamma^\mu \psi_\mu^m$. After an obvious field redefinition, 
\begin{equation}
	\psi_\mu^i \to \psi_\mu^i + \frac{1}{18\sqrt2}\, A^{-1\, ij} A_j{}^{klm}\, \gamma_\mu \chi_{klm} \, , 
\end{equation} 
we can get rid of the mixed term, so that the spin-1/2 mass matrix becomes
\begin{equation}
	\label{spin12}
	{\cal M}_{(1/2)}{}^{ijk,lmn} =	  
	\frac{\sqrt{2}}{12}\,
	  \epsilon^{ijkpqr[lm}\,A^{n]}{}_{pqr} + \frac{\sqrt2}{9} A_a^{ijk} A^{-1\, ab} A_{b}^{lmn}.
\end{equation}
The goldstinos will be null eigenvectors of this matrix, hence they will not contribute to the supertrace formulas.

The mass matrix (\ref{spin12}) contains the inverse of the gravitino shift and this is an unpleasant feature that complicates the computation of the traces. For this reason we will now argue that in order to compute the traces of its even powers it is sufficient to use the original mass matrix $M^{ijk,lmn}$, provided we subtract from the final result a suitable factor proportional to the trace of the corresponding power of the gravitino mass matrix. The argument uses the fact that the mass of the goldstino is really a gauge artifact and that the term we added to the original spin-1/2 mass matrix is fully projected in the goldstino directions to make their mass vanish.

\subsubsection{More on the spin 1/2 mass matrix} 
\label{sub:More on the spin 1/2 mass matrix}

When supersymmetry is fully broken we have 8 goldstino fields, determined by linear combinations of the spin 1/2 fields projected along the $A_2$ direction: $\eta^i \sim A^i{}_{jkl} \,\chi^{jkl}$.
These 8 directions among the 56 described by the spin 1/2 fields are eigenvectors with non-vanishing eigenvalue of $M$ as well as of the shift term, but they are null eigenvectors of the redefined mass matrix ${\cal M}_{(1/2)}$. This can be seen explicitly by making use of the critical point condition (\ref{vacuum}) and of a few identities following from the so-called T-tensor identities of $N=8$ supergravity, or equivalently from the quadratic constraint on the embedding tensor $\Theta$, which is reported in the Appendix.

First, by using the Takagi factorization theorem, we can diagonalize the gravitino mass matrix by means of a unitary matrix $U$, so that $A_1 = U^T D U$ and the diagonal matrix $D$ contains the gravitino masses $m_{\alpha} >0$, where $\alpha$ runs over the 8 eigenvalues. We can then construct the eigenvectors of $A_1$ by applying the same unitary matrix to the basis of (real) eigenvectors of $D$, $e^i_{\alpha} = \delta^i_{\alpha}$. Indeed the vectors $V_{\alpha} = U^{-1} e_{\alpha}$ satisfy
\begin{equation}
	\sqrt2\, A_{ij} V_\alpha^j = m_{\alpha} V_{\alpha\, i},
	\label{gravitinoeig}
\end{equation}
where there is no sum on the right hand side and $V_{\alpha\, i} = (V^i_{\alpha})^*$.
As follows from a direct inspection of (\ref{Lmass}), the corresponding goldstino directions are given by $A^i{}_{jkl} V_{\alpha\, i}$.
In fact, we can now check that the goldstinos are eigenvectors of the original spin-1/2 mass matrix with eigenvalue $-2 \,m_{\alpha}$, where $m_{\alpha}$ is the corresponding gravitino mass.
This is done by using the identity (\ref{finalidentity}) in the explicit multiplication of the mass matrix with the goldstino vector as follows:
\begin{equation}
	M^{ijk,pqr} A^s{}_{pqr} V_{\alpha s} = \frac{\sqrt2}{12}\,\epsilon^{ijkuvwpq} A^r{}_{uvw} A^s{}_{pqr} V_{\alpha s} \stackrel{(\ref{finalidentity})}{=}
	-2\sqrt2 \, A_p{}^{ijk} A^{ps} V_{\alpha s} \stackrel{(\ref{gravitinoeig})}{=} -2 \,m_{\alpha}\,  A_p{}^{ijk} V^p_\alpha.
\end{equation}
Also, using now the supersymmetric Ward identity (\ref{piidentity}) and the fact that we are analyzing vacua with vanishing cosmological constant, we recover the fact that the goldstino is a null eigenvector of ${\cal M}_{(1/2)}$.
In detail:
\begin{eqnarray}
	\left({\cal M}_{(1/2)} - M\right)^{ijk,pqr} A^s{}_{pqr} V_{\alpha \,s} & = & \frac{\sqrt2}{9} A_m^{ijk} A^{-1\, mn} A_{n}^{pqr} A^s{}_{pqr} V_{\alpha \, s} 
\nonumber \\ &
\stackrel{(\ref{piidentity2})}{=} & 2\sqrt2 \, A_p{}^{ijk} A^{ps} V_{\alpha\,s} \stackrel{(\ref{gravitinoeig})}{=} 2 \,m_{\alpha}\,  A_p{}^{ijk} V^p_{\alpha} \, .
\end{eqnarray}
By construction $\left({\cal M}_{(1/2)} - M\right)^{ijk,pqr}$ is fully projected on the goldstino directions and therefore any other orthogonal direction will be a null eigenvector for such matrix.
This implies that if we are interested in computing the sum of the eigenvalues of $[{\cal M}_{(1/2)}^2]^{k}$, we can actually first compute the sum of the eigenvalues of $(M^2)^{k}$ and then remove from the result $2^{2k}$ times the sum of the eigenvalues of the gravitino mass to the same power.
Since the matrix $M$ has a simpler form that ${\cal M}_{(1/2)}$, we will follow this approach in the next section, where we will compute the supertrace formulas using 
\begin{equation}
	{\rm Tr} \, {\cal M}_{(1/2)}^2 = {\rm Tr}\, M^2 - 4 \,{\rm Tr} \, {\cal M}_{(3/2)}^2
	\label{quadraticchi}
\end{equation}
and
\begin{equation}
	{\rm Tr} \, {\cal M}_{(1/2)}^4 = {\rm Tr}\, M^4 - 16\, {\rm Tr} \, {\cal M}_{(3/2)}^4.
	\label{quarticchi}
\end{equation}




\section{Supertrace mass formulas} 
\label{sec:The supertrace mass formula}

In this section we evaluate the supertrace mass formulas of Eq.~(\ref{Supertrace}) for $k=1$ and $k=2$, and show that they both vanish for any Minkowski vacuum, as in Eq.~(\ref{result1}). Our procedure is rather straightforward. First we compute the trace of the $2k$-th power of the spin-$J$ mass matrices described in the previous section (i.e. Tr\,$[{\cal M}_{(J)}^2]^k$). These traces are SU(8) singlets constructed in terms of the $A_1$ and $A_2$ tensors introduced above. Hence they are given by linear combinations of quadratic or quartic contractions of these same tensors, for $k=1$ or $k=2$ respectively.
We then simplify them by using the quadratic identities in the $\mathbf{63}$ of SU(8), which allow us to replace most of the terms quadratic in $A_2$ by quadratic contractions of $A_1$. Finally, we compute (\ref{Supertrace}) and show that for $k=1$ it vanishes identically, while for $k=2$ we prove that it is equivalent to a linear combination of terms that are also vanishing upon using the quadratic constraints and the critical point condition.

For more details on the derivation of the identities following from the quadratic constraints we refer the reader to the Appendix.

\subsection{The quadratic supertrace} 
\label{sub:The quadratic supertrace}

As explained in the introduction to this section, the first step is to compute the traces of the squared mass matrices of the various fields.
Some of them follow trivially from their definitions, such as the trace of the square mass matrix of the gravitinos
\begin{equation}
	{\rm Tr} \, {\cal M}_{(3/2)}^2 = 2 \,A_{ij} A^{ij} \, ,
\end{equation}
the trace of the vector mass matrix
\begin{equation}
	{\rm Tr} \,{\cal M}_{(1)}^2 = {\cal M}_{ij}{}^{ij} + {\cal M}^{ij}{}_{ij} = \frac23 \,A_{i}{}^{jkl} A^{i}{}_{jkl} \stackrel{\Lambda = 0}{=}  12 \,A_{ij}A^{ij} \, , 
\end{equation}
and the trace of the scalar mass matrix:
\begin{equation}
	{\rm Tr}\, {\cal M}_{(0)}^2 = {\cal M}_{ijkl}{}^{ijkl}=\frac72 \, A_{i}{}^{jkl} A^{i}{}_{jkl} - 35\, A_{ij}A^{ij} \stackrel{\Lambda = 0}{=} 28\,A_{ij}A^{ij} \, .
\end{equation}
As we can see all of them are proportional to each other by using the condition that the vacuum has a vanishing cosmological constant.
For the spin-1/2 fermions the computation is a bit more delicate, but we can use the technique explained in the previous section to simplify the procedure.
By using (\ref{quadraticchi})
\begin{equation}
	{\rm Tr}\, {\cal M}_{(1/2)}^2 = {\rm Tr} \, M^2 - 8 \, A_{ij} A^{ij}
\label{trmma}
\end{equation}
and, by direct computation,
\begin{equation}
	 {\rm Tr}\, M^2  = M^{ijk,pqr} M_{ijk, pqr} = 2 \, A_{i}{}^{jkl} A^{i}{}_{jkl}.
\end{equation}
This means that also this trace is proportional to the trace of the square of the gravitino mass matrix on a Minkowski vacuum:
\begin{equation}
	{\rm Tr}\, {\cal M}_{(1/2)}^2 = 2 \, A_{i}{}^{jkl} A^{i}{}_{jkl}- 8 \,A_{ij} A^{ij} \stackrel{\Lambda = 0}{=} 28 \,A_{ij} A^{ij}.
\label{trmaa}
\end{equation}
As explained in the previous section, the derivation of Eqs.~(\ref{trmma}) and (\ref{trmaa}) required the use of the critical point condition of Eq.~(\ref{vacuum}). Inserting these expressions in the supertrace formula we then see that the coefficients conspire to make the final expression identically zero:
\begin{equation}
	\begin{array}{rccccccl}
	{\rm Str}{\cal M}^2  &=& &-4\, {\rm Tr}\, {\cal M}_{(3/2)}^2 &+ 3\, {\rm Tr}\, {\cal M}_{(1)}^2 &-2\, {\rm Tr} \,{\cal M}_{(1/2)}^2& + {\rm Tr} \,{\cal M}_{(0)}^2 &\\[3mm]
	&=& (&-8& + 36 &- 56& + 28&)A_{ij}A^{ij} = 0.
	\end{array}
	\label{supertracefinal}
\end{equation}


\subsection{The quartic supertrace} 
\label{sec:The quartic supertrace} 

The computation of the trace of the quartic mass matrix is again straightforward for the gravitinos:
\begin{equation}
	{\rm Tr}\,{\cal M}_{(3/2)}^4 = 4 \,A_{ij}A^{jk} A_{kl} A^{li}.
\end{equation}
On the other hand, the expressions for the other fields are now more complicated and, for some of the simplifications, we made use of a symbolic computer algebra software \cite{Peeters:2006kp,Peeters:2007wn}.
We give here the value of the traces, where a first simplification has been made by using the quadratic identities in the $\mathbf{63}$ of SU(8).
For the vector fields we obtain
\begin{equation}
	\begin{array}{rcl}
	{\rm Tr}\,{\cal M}_{(1)}^4 &=&\displaystyle - \frac{5}{2}\, {A}_{a b} {A}_{c d} {A}^{a b} {A}^{c d} - 16\, {A}_{a b} {A}_{c d} {A}^{a c} {A}^{b d} + \frac{3}{8}\, {A}_{a}\,^{b c d} {A}_{b}\,^{a e f} {A}^{g}\,_{c d i} {A}^{i}\,_{e f g} \\[3mm]
	&&-\displaystyle  \frac{1}{2}\, {A}_{a}\,^{b c d} {A}_{b}\,^{e f g} {A}^{a}\,_{e f i} {A}^{i}\,_{c d g} + \frac{3}{8}\, {A}_{a}\,^{b c d} {A}_{e}\,^{f g i} {A}^{a}\,_{b f g} {A}^{e}\,_{c d i}
	\\[3mm]
	&& \displaystyle + \frac{1}{2}\, {A}_{a}\,^{b c d} {A}_{b}\,^{e f g} {A}^{a}\,_{c e i} {A}^{i}\,_{d f g} 
 - \frac{1}{2}\, {A}_{a}\,^{b c d} {A}_{b}\,^{a e f} {A}^{g}\,_{c e i} {A}^{i}\,_{d f g}.	
	\end{array}
\end{equation}
For the spin 1/2 fields, the result follows from (\ref{quarticchi})
\begin{equation}
	{\rm Tr}\,{\cal M}_{(1/2)}^4 = 	{\rm Tr}\, {M}^4 - 64\, A_{ij}A^{jk} A_{kl} A^{li},
\end{equation}
where the first term is
\begin{equation}
	{\rm Tr}\,{M}^4 = 4 A_{a}{}^{b c d} A_{b}{}^{a e f} A^{g}{}_{c e i} A^{i}{}_{d f g} - 
8 A_{a}{}^{b c d} A_{b}{}^{e f g} A^{a}{}_{c e i} A^{i}{}_{d f g} + 
8 A_{a b} A_{c d} A^{a c} A^{b d} + 80 A_{a b} A_{c d} A^{a b} A^{c d}.
\end{equation}
Finally, for the scalar fields we have
\begin{equation}
	\begin{array}{rcl}
	{\rm Tr}\,{\cal M}_{(0)}^4 &=& \displaystyle - 96 {A}_{a b} {A}_{c d} {A}^{a c} {A}^{b d} - \frac{13}{2} {A}_{a b} {A}_{c d} {A}^{a b} {A}^{c d}  \\[3mm]
	&&\displaystyle + \frac{11}{8}\, {A}_{a}{}^{b c d} {A}_{b}{}^{a e f} {A}^{g}{}_{c d i} {A}^{i}{}_{e f g} - \frac{5}{2} {A}_{a}{}^{b c d} {A}_{b}{}^{e f g} {A}^{a}{}_{e f i} {A}^{i}{}_{c d g} \\[3mm]
	&& \displaystyle + \frac{11}{8} {A}_{a}{}^{b c d} {A}_{e}{}^{f g i} {A}^{a}{}_{b f g} {A}^{e}{}_{c d i} + \frac{5}{2}\, {A}_{a}{}^{b c d} {A}_{b}{}^{e f g} {A}^{a}{}_{c e i} {A}^{i}{}_{d f g} - \frac{1}{2}\, {A}_{a}{}^{b c d} {A}_{b}{}^{a e f} {A}^{g}{}_{c e i} {A}^{i}{}_{d f g}.
	\end{array}
\end{equation}
Altogether, the quartic supertrace formula is given by the combination
\begin{equation}
	\begin{array}{rcl}
	{\rm Str}{\cal M}^4 &=& \displaystyle - 48\, {A}_{a b} {A}_{c d} {A}^{a c} {A}^{b d} - 174\, {A}_{a b} {A}_{c d} {A}^{a b} {A}^{c d} \\[3mm]
	&& \displaystyle + \frac{5}{2}\, {A}_{a}\,^{b c d} {A}_{b}\,^{a e f} {A}^{g}\,_{c d i} {A}^{i}\,_{e f g} - 4\, {A}_{a}\,^{b c d} {A}_{b}\,^{e f g} {A}^{a}\,_{e f i} {A}^{i}\,_{c d g} \\[3mm]
	&& \displaystyle + \frac{5}{2}\, {A}_{a}\,^{b c d} {A}_{e}\,^{f g i} {A}^{a}\,_{b f g} {A}^{e}\,_{c d i} + 20\, {A}_{a}\,^{b c d} {A}_{b}\,^{e f g} {A}^{a}\,_{c e i} {A}^{i}\,_{d f g} \\[3mm]
	&& - 10\, {A}_{a}\,^{b c d} {A}_{b}\,^{a e f} {A}^{g}\,_{c e i} {A}^{i}\,_{d f g},
	\end{array}
\end{equation}
which can be rewritten as a linear combination of 4 different quartic terms that vanish identically because of the quadratic constraints evaluated at Minkowski critical points of the scalar potential:
\begin{equation}
	{\rm Str}{\cal M}^4 = 	-12\, I_1 - 5 \, I_6 + 16 \, I_7+\frac52 \, I_5 = 0.
\end{equation}
The detailed expression for the identically vanishing terms $I_1, \ldots , I_7$  can be found in the Appendix, where we also discuss their derivation.




\section{Minkowski vacua with unbroken U(1) factors} 
\label{sec:Minkowski vacua with unbroken U(1) factors}

After showing that ${\rm Str} \, {\cal M}^2 = {\rm Str} \, {\cal M}^4 =0$ at any classical Minkowski vacuum of an arbitrary gauged $N=8$ supergravity, we would like to go further and derive some results also on ${\rm Str} \, {\cal M}^6$, ${\rm Str} \, {\cal M}^8$ and on the general properties of the one-loop effective potential $V_1$. To do so, the strategy adopted in Section~3 does not look powerful enough. We then proceed by making two assumptions, which are satisfied by all presently known explicit gaugings of $N=8$ supergravity that lead to classical Minkowski vacua with fully broken supersymmetry. Our first assumption is that the subgroup of the gauged symmetry left unbroken by the vacuum contains at least a U(1) factor, under which the eight supercharges $Q_i$ ($i=1,\ldots,8$) transform non-trivially. The unbroken gauge group must be a subgroup of SU(8), thus we can expect in general several U(1)$_A$ factors ($A=1,\ldots,n$; $n \le 7$). We will denote by $\vec{q}_i \equiv (q_i^1,\ldots,q_i^n)$ the vector of the charges of $Q_i$ with respect to U(1)$^n$. We recall that the 8 gravitinos of helicity $\pm 3/2$, the 28 gauge bosons of helicity $\pm1$, the 56 fermions of helicity $\pm 1/2$ (including the goldstinos) and the 70 scalars of helicity 0 (including the Goldstone bosons) are obtained by acting repeatedly with the supercharges on the graviton of helicity $+ 2$. This uniquely determines the charges of all the states, according to the following scheme:
\begin{eqnarray}
| +2 \rangle : & & \vec{0} \, , 
\nonumber \\
| +3/2 \, , \, i \rangle =  Q_i \, | +2 \rangle : & & \vec{q}_i \, , 
\nonumber \\
| +1 \, , \, [ij] \rangle =  Q_i \, Q_j | +2 \rangle : & & \vec{q}_i + \vec{q}_j \, , 
\nonumber \\
| +1/2 \, , \, [ijk] \rangle =  Q_i \, Q_j \, Q_k | +2 \rangle : & & \vec{q}_i + \vec{q}_j + \vec{q}_k \, , 
\nonumber \\
| 0 \, , \, [ijkl] \rangle =  Q_i \, Q_j \, Q_k \, Q_l | +2 \rangle : & & \vec{q}_i + \vec{q}_j + \vec{q}_k + \vec{q}_l \, , 
\label{U1charges}
\end{eqnarray}
and similarly for the CPT-conjugates. Acting with all 8 supercharges on the graviton of helicity $+2$ gives back the graviton of helicity $-2$, which leads to the constraint:
\begin{equation}
\sum_{i=1}^8 \vec{q}_i = \vec{0} \, . 
\label{CPT}
\end{equation}
Our second assumption is that the spectrum can be represented as follows:
\begin{eqnarray}
|2 \rangle : & & M^2 = 0 \, , 
\nonumber \\
|3/2 \, , \, i \rangle : & & M_i^2 = \left( \vec{q}_i \right)^2 \, , 
\nonumber \\
|1 \, , \, [ij] \rangle : & & M_{ij}^2 = \left( \vec{q}_i + \vec{q}_j \right)^2  \, , 
\nonumber \\
|1/2 \, , \, [ijk] \rangle : & & M_{ijk}^2 = \left( \vec{q}_i + \vec{q}_j + \vec{q}_k \right)^2  \, , 
\nonumber \\
| 0 \, , \, [ijkl] \rangle : & & M_{ijkl}^2 = \left( \vec{q}_i + \vec{q}_j + \vec{q}_k + \vec{q}_l \right)^2 \, , 
\label{U1spectrum}
\end{eqnarray}
where the scalar products of the charge vectors must be taken with a suitable field-dependent real diagonal metric, not necessarily positive definite (having it positive definite, however, guarantees the absence of tachyons for the corresponding field configurations):
\begin{equation}
\vec{q}_i \cdot \vec{q}_j = \sum_{A=1}^n q_i^A \, q_j^A \ \mu_A^2 \, .
\label{U1metric}
\end{equation}
Implicitly, we are working in a gauge where the goldstinos, which provide the $\pm 1/2$ helicity components of the massive gravitinos, are assigned masses equal to the corresponding gravitino masses, and the Goldstone bosons, which provide the longitudinal component of the massive vectors, are assigned masses equal to the corresponding vector boson masses. This simplifies the computation of the supertraces, since we can now sum also over unphysical states (goldstinos and Goldstone bosons) and write:
\begin{equation}
{\rm Str} \, {\cal M}^{2k} =
{\rm Tr} \, [{\cal M}_{(0)}^2]^k - 2 \ 	{\rm Tr} \, [{\cal M}_{(1/2)}^2]^k + 2 \ {\rm Tr} \, [{\cal M}_{(1)}^2]^k 
- 2 \ {\rm Tr} \, [{\cal M}_{(3/2)}^2]^k \, .
\label{Supertracerev}
\end{equation}
It is immediate to check that, under the two assumptions specified above,
\begin{equation}
{\rm Str} \, {\cal M}^2 = {\rm Str} \, {\cal M}^4 = {\rm Str} \, {\cal M}^6 =  0 \, , 
\label{m6zer}
\end{equation}
and
\begin{equation}
{\rm Str} \, {\cal M}^8 =  40320 \, \sum_{A=1}^n \, \left( \prod_{i=1}^8 q_i^A \right) \, \mu_A^8 \, .
\end{equation} 

We now show that, under our assumptions, there are always four unbroken U(1) factors ($n=4$) and the 8 supercharges always come in pairs, with opposite charge vectors ($\vec{q}_1 = - \vec{q}_2$, $\vec{q}_3 = - \vec{q}_4$, $\vec{q}_5 = - \vec{q}_6$, $\vec{q}_7 = - \vec{q}_8$).  Incidentally, this explains why all the presently known models of this kind have at most 4 independent gravitino mass parameters, corresponding to `Dirac' mass terms rather than `Majorana' ones. The proof goes as follows. Since, by assumption, there is at least one unbroken U(1), there must be an associated massless vector, i.e., according to Eq.~({\ref{U1spectrum}), there must be a couple $[ij]$, say $[12]$, for which $\vec{q}_1 = - \vec{q}_2$. If supersymmetry is fully broken on the Minkowski vacuum, there must also be 8 goldstinos with the same charge vectors of the corresponding gravitinos. In other words, according to Eq.~(\ref{U1charges}), for each value of $i$ it must be $\vec{q}_i = \vec{q}_i + \vec{q}_j + \vec{q}_k$ for a suitable couple $[jk]$. We know already that  $\vec{q}_1 = - \vec{q}_2$, thus for $i=3,4,5,6,7,8$ we can use $[jk] = [12]$. On the other hand, for $i = 1$ and $i = 2$ we must use some $[jk] \ne [12]$. The only possibility is to have two other charge vectors opposite to each other, say $\vec{q}_3 = - \vec{q}_4$. Now we can look again at the vector boson spectrum and see that, without additional assumptions on the charges, there are in principle 2 massless vectors, corresponding to $[ij] = [12], \, [34]$, and 26 massive ones. Correspondingly, there should be 26 Goldstone bosons $[ijkl]$, with $[ij]$ different from the two combinations just mentioned and $\vec{q}_k = - \vec{q}_l$. If instead we accept that $\vec{q}_1 = - \vec{q}_2 = \vec{q}_3 = - \vec{q}_4$, then we have 4 massless vectors, corresponding to $[ij] = [12], \, [34], \, [14], \, [23]$, and 24 massive ones. Correspondingly, there should be 24 Goldstone bosons $[ijkl]$, with $[ij]$ different from the four combinations just mentioned and $\vec{q}_k = - \vec{q}_l$. In any case, for at least a subset of $[ij] = [13], \, [24], \, [14], \, [23]$ we have a problem, and we must impose another relation between charge vectors, for example $\vec{q}_5 = - \vec{q}_6$. At this point the CPT constraint of Eq.~(\ref{CPT}) imposes the additional relation $\vec{q}_7 = - \vec{q}_8$. In conclusion, there are at least 4 unbroken U(1) factors and at most 4 independent charge vectors $\vec{q}_i$. This means that, if there are more than 4 unbroken U(1) factors on the vacuum, we can always redefine these U(1) factors in such a way that at most 4 of them admit charged states, while the remaining ones are `spectators', without charged states. It also means that, for the U(1) factors that admit charged states,
\begin{equation}
\prod_{i=1}^8 q_i^A = 
\prod_{i=even}^8 \left( q_i^A \right)^2 = 
\prod_{i=odd}^8 \left( q_i^A \right)^2 > 0 \, ,
\end{equation} 
thus
\begin{equation}
{\rm Str} \, {\cal M}^8 > 0 \, .
\label{m8pos}
\end{equation} 
{}From the existence of at least 4 unbroken U(1) factors, corresponding to $[ij]$ = [12],  [34], [56],  [78], we can also deduce that there are at least 6 massless scalars in the physical spectrum, corresponding to the 6 independent combinations $[ijkl]$ with $[ij], \, [kl] = [12], \, [34], \, [56], \, [78]$.

The result of Eq.~(\ref{m8pos}) is very important because, when combined with the one in Eq.~(\ref{m6zer}), it is strongly correlated with the sign of the one-loop effective potential $V_1$, as we now discuss. If the squared mass eigenvalues $M_a^2$ had small fluctuations $\delta M_a^2 \equiv M_a^2 - \langle M^2 \rangle$ around the average $\langle M^2 \rangle$, computed over all bosonic and fermionic degrees of freedom, including the gravitons, it would be immediate to see that
\begin{equation}
{\rm Str} \, {\cal M}^{2k} = 0 \quad (k=0,\ldots,n) 
\quad \Rightarrow \quad
{\rm Str} \, (\delta M^2)^{k} = 0  \quad (k=0,\ldots,n)  \, .
\end{equation}
In particular, if ${\rm Str} \, {\cal M}^2 = {\rm Str} \, {\cal M}^4 = {\rm Str} \, {\cal M}^6 =  0$, then ${\rm Str} \, {\cal M}^8 =  {\rm Str} \, (\delta M^2)^{4}$, and we could write
\begin{equation}
V_1 =  \frac{1}{64 \, \pi^2} \ {\rm Str} \,   \left(  -  \frac{{\cal M}^8}{2 \, \langle M^2 \rangle^2} + \ldots \right) \, . 
\end{equation}
Regrettably, $| \delta M_a^2 / \langle M^2 \rangle |$ is of order one for the class of models being considered in this section. In this case, a possible strategy could start from considering the function
\begin{equation}
F (x^2) = \frac{1}{64 \, \pi^2} \ {\rm Str} \,  [ {\cal M}^4 \ \log ({\cal M}^2 + x^2) ] \, .
\end{equation}
Its properties are that $F(0) = V_1$ and that, for $x^2$ much larger than all squared-mass eigenvalues, and assuming ${\rm Str} \, {\cal M}^0 = {\rm Str} \, {\cal M}^2 = {\rm Str} \, {\cal M}^4 =  0$,
\begin{equation}
F(x^2) \rightarrow  \frac{1}{64 \, \pi^2} \ {\rm Str} \,   \left( \frac{{\cal M}^6}{x^2}  -  \frac{{\cal M}^8}{2 \, x^4} + \ldots \right) \, . 
\end{equation}
Moreover,
\begin{equation}
\frac{dF}{dx^2} =  \frac{1}{64 \, \pi^2} \ {\rm Str} \,   \left(  {\cal M}^4 \  \frac{1}{{\cal M}^2 + x^2} \right)  = \frac{1}{64 \, \pi^2} \ {\rm Str} \,   \left(   \frac{x^4}{{\cal M}^2 + x^2} \right)  
\end{equation}
goes to zero both for $x^2 \rightarrow 0$ and for $x^2 \rightarrow \infty$, and for large $x^2$ goes as
\begin{equation}
\frac{dF}{dx^2} \rightarrow  \frac{1}{64 \, \pi^2} \ {\rm Str} \,   \left( - \frac{ {\cal M}^6}{x^4}  +   \frac{{\cal M}^8}{x^6}  + \ldots \right)  \, .
\end{equation}
In the class of gaugings of this section, ${\rm Str}\, {\cal M}^6 = 0$ and ${\rm Str}\, {\cal M}^8 > 0$, thus for large $x^2$ it is $F<0$ and $dF/dx^2 > 0$. It would be enough to show that $dF/dx^2$ has no other zeros at finite values of $x^2$ to prove the desired correlation. This line of reasoning is empirically confirmed by numerical studies of all the models considered later in this section, and of a model with the same spectrum of the $N=4$ truncation of the $N=8$ CSS gauging \cite{D'Auria:2003jk}. The numerical study of a $N=6$ theory where \cite{Andrianopoli:2002rm, Villadoro:2004ec} ${\rm Str} \,  {\cal M}^2 = {\rm Str} \,  {\cal M}^4 = 0$ but ${\rm Str} \,  {\cal M}^6 < 0$ also gives $V_1 < 0$, as expected. 

In the following subsections we show how all the presently known explicit gaugings of $N=8$ supergravity that lead to classical Minkowski vacua with fully broken supersymmetry fulfill the two assumptions described above, with suitably defined charge vectors $\vec{q}_i$ and U(1) metrics $\mu_A^2$, and discuss their properties in detail. We start by recalling some known properties of the CSS gauging \cite{Cremmer:1979uq}, then we spell out the properties of the gaugings recently discussed in \cite{DallAgata:2011aa}, pointing out similarities and differences. As already anticipated above, we also check numerically that for all these gaugings the result of Eq.~(\ref{result4}) holds.

\subsection{The CSS gauging and the role of USp(8)} 
\label{subsec:CSS gauging}

The best studied example of spontaneously broken $N=8$, $D=4$ supergravity on a classically flat background is the four-parameter CSS gauging \cite{Cremmer:1979uq}, originally obtained via generalized dimensional reduction from five-dimensional maximal supergravity. Such gauging includes the three-parameter gauging previously obtained by Scherk and Schwarz \cite{Scherk:1979zr} by generalized dimensional reduction of eleven-dimensional supergravity, as well as other flat gaugings that can be obtained from such a theory or from the type-II ten-dimensional supergravitites in the presence of general geometric fluxes \cite{D'Auria:2005er, D'Auria:2005dd, D'Auria:2005rv}. It was realized already in the original CSS paper, and explained in group-theoretical terms in \cite{Ferrara:1979fu}, that for the CSS gauging ${\rm Str} \, {\cal M}^2 = {\rm Str} \, {\cal M}^4 = {\rm Str} \, {\cal M}^6 =  0$, but ${\rm Str} \, {\cal M}^8 \ne 0$ if there is no residual supersymmetry on the vacuum.  A more detailed analysis of the one-loop finiteness was carried out in  \cite{Sezgin:1981ac, Sezgin:1982pk}, where it was also realized that at the origin of moduli space the one-loop contribution to the vacuum energy density is negative definite. It is also known (see e.g. \cite{Andrianopoli:2002vy}) that the classical Minkowski vacua of the CSS gauging have three complex flat directions, parameterizing the coset $[SU(1,1)/U(1)]^3$.

In this subsection we briefly recall the ingredients that enter the calculation of the CSS spectrum, showing how they fit in the more general framework introduced at the beginning of this section. We also go further and quote the results of a numerical study  \cite{CatDalInvZwi}, which shows that for the CSS gauging all the degenerate classical Minkowski vacua with fully broken supersymmetry are destabilized by the one-loop corrections: $V_1$ is negative definite and has no critical points within the classical moduli space. This leads to a dead end, at least for those values of the parameters and background fields for which the perturbative expansion makes sense.

Generically, the CSS gauging \cite{Cremmer:1979uq} promotes U(1)$\ltimes T^{24}$ to a local symmetry, where $T^{24}$ denotes the number of translations along non-compact directions. In addition, the remaining 3 vectors correspond to `spectator' U(1) factors, in the sense there are no fields carrying non-vanishing charges with respect to them. We then fall in the formalism described at the beginning of this section. The classical potential is positive semi-definite and has only Minkowski vacua. The spectrum  is described by four real positive parameters. They are denoted here by $(e_1, e_2, e_3, e_4)$, rather than $(m_1, m_2, m_3, m_4)$ as in the original literature, to stress the fact that they are the absolute values of the charges $q_i$ of the 8 supersymmetry generators $Q_i$, with respect to the unbroken U(1) factor in the gauge group [U(1) $\subset$ USp(8) $\subset$ SU(8) $\subset E_{7(7)}$]:
\begin{equation}
q_1 = - q_2 = e_1 \, , 
\quad
q_3 = - q_4 = e_2 \, , 
\quad
q_5 = - q_6 = e_3 \, , 
\quad
q_7 = - q_ 8 = e_4 \, .  
\label{csscharges}
\end{equation}
For the CSS gauging, the fact that the charges are paired in 4 couples is a direct consequence of the canonical form of a single U(1) generator in the fundamental representation of USp(8). We do not display here the charges of the supersymmetry generators with respect to the other 3 `spectator' U(1)s, since they are identically vanishing. The spectrum of the CSS gauging is described by the general formulas of Eqs.~(\ref{U1spectrum}) and  (\ref{U1metric}), with
\begin{equation}
\mu^2 =  \phi^2 \, ,
\label{massformcss}
\end{equation}
where $\phi >0 $ is one of the six real flat directions of the classical potential, which provides a universal scaling factor for all mass terms of the theory and in which we can reabsorb the gauge coupling constant. If all the $e_a$ ($a=1,2,3,4$) are non-vanishing, then supersymmetry is fully broken on the vacuum, and 8 would-be goldstinos are removed from the spectrum. There are always at least four massless vectors, associated with the U(1) in  the gauged U(1)$\ltimes T^{24}$ and the 3 additional `spectator' U(1) factors. Correspondingly, at most 24 would-be-Goldstone bosons are removed from the spectrum. In the scalar sector, there are always at least six massless real degrees of freedom, corresponding to at least three complex flat directions of the classical potential. For special values of the parameters the unbroken gauge group can be enhanced, with the associated vectors becoming massless and the associated would-be Goldstone bosons moving back to the physical spectrum at zero mass. Also, for special values of the parameters there can be additional classical flat directions.  The case with the most degenerate spectrum is obtained for $e_1=e_2=e_3=e_4$. In such a case, the gauged group is just U(1)$\ltimes T^{12} \simeq$~CSO(2,0,6), and all massive states have squared masses that are integer multiples of a universal gravitino mass $m_{3/2}^2$: $4 \, m_{3/2}^2$ for 12 massive vectors; 
$9 \, m_{3/2}^2$ and $m_{3/2}^2$ for 8 and 40 massive spin-1/2 fermions, respectively; $4 \, m_{3/2}^2$ and $16 \, m_{3/2}^2$ for 20 and 2 massive scalars, respectively.

We recall here the role of USp(8) for the vanishing of the supertraces in the case of the CSS gauging. For CSS, all the mass matrices can be written as matrices representing the generator of a single U(1) $\subset $ USp(8), in different representations. The moduli dependence factorizes and all mass matrices can be rewritten in terms of the one in the fundamental representation, so that the supertrace relations follow from USp(8) algebraic properties \cite{Ferrara:1979fu}. We stress that such an argument cannot be applied to generic gaugings, because the residual gauge group on a Minkowski vacuum is not necessarily a subgroup of USp(8) and, even if that were the case, the mass matrices may represent multiple independent U(1) factors, weighted by different functions of the moduli.

On the basis of the above results, the full field-dependence of the one-loop effective potential along the three complex flat directions of the classical potential is summarized by:
\begin{equation}
V_{1}  = \frac{\phi^4}{64 \, \pi^2} \ {\sum_a}^\prime (2 J_a + 1) (-1)^{2 J_a} \ (q_a)^4 \, \log (q_a)^2 \, , 
\end{equation}
where the prime means that the sum is taken only for those values of $a = i, \, [ij], \, [ijk], \, [ijkl]$ for which $q_a \ne 0$, and $q_{[ij]}=q_i + q_j$, $q_{[ijk]}=q_i + q_j + q_k$, $q_{[ijkl]}=q_i + q_j + q_k + q_l$ [see Eq.~(\ref{U1charges})]. Besides vanishing, as expected, in the $N \ge 2$ supersymmetric limit in which any of its four arguments goes to zero, the one-loop effective potential $V_1$ is negative semi-definite, and vanishes {\em only} in the supersymmetric limit discussed above. In the present case of large fluctuations in the mass spectrum, for which we do not have a full analytical proof, this is confirmed by a numerical study \cite{CatDalInvZwi}. This leads to a dead end for the CSS gauging, since all its degenerate classical Minkowski vacua with fully broken supersymmetry are destabilized by the one-loop corrections: $V_1$ has no critical points within the classical moduli space. 


\subsection{Other gaugings with G ${\mathbf \subset}$ SL(8,${\mathbf \mathbb R}$) } 
\label{subsec:DIgaugings}

In this subsection we study in some detail a broad class of $N=8$ gaugings recently discussed in \cite{DallAgata:2011aa}, which lead to classical Minkowski vacua with fully broken supersymmetry. We show that they all fulfill the assumptions spelled out at the beginning of this section, with the following charge vector assignments:
\begin{eqnarray}
\vec{q}_1 = - \vec{q}_2 = (+1,+1,+1,+1) \, , 
& & 
\vec{q}_3 = - \vec{q}_4 = (+1,+1,-1,-1) \, , 
\nonumber \\
\vec{q}_5 = - \vec{q}_6 = (+1,-1,+1,-1) \, , 
& & 
\vec{q}_7 = - \vec{q}_8 = (+1,-1,-1,+1) \, ,
\label{so62charges}
\end{eqnarray}
and the metric of Eq.~(\ref{U1metric}) specified by mass parameters $\mu_A^2$ that depend on the specific gaugings and parameter choices. We point out similarities and differences with the CSS gauging. We describe the spectrum in detail and find that, in contrast with the CSS gauging, for generic values of the parameters there can be classical instabilities corresponding to tachyonic masses for some of the scalar fields. We study in detail the one-loop effective potential for those parameter choices that are free from classical tachyonic instabilities and find numerically that, as for the CSS gauging, $V_1 < 0$ at every point of the classical moduli space.  As for the CSS gauging, we conclude that there is no locally stable one-loop vacuum with broken supersymmetry and positive or vanishing vacuum energy. In contrast with the CSS gauging, however, we cannot exclude the presence of locally stable AdS vacua at the one-loop level.

The crucial technical tool of Ref.~\cite{DallAgata:2011aa} is to use the $E_{7(7)}$ duality group of the ungauged theory, which acts both on the embedding tensor and on the scalar fields parameterizing the $E_{7(7)}/SU(8)$ coset space, to map any critical point of the scalar potential to the base point of the scalar manifold. This means that we can survey the different vacua of the gauged theory by staying always at the origin in field space, and studying the dependence of the embedding tensor on its parameters. For the purposes of the present paper, we shall restrict our attention to forms of the embedding tensor that are compatible with Minkowski vacua with fully broken supersymmetry. An exhaustive classification is not available, thus we will consider only the explicit examples given in \cite{DallAgata:2011aa}, whose common feature is to have a gauged group G~$\subset$~SL(8,${\mathbb R}$). The embedding  tensor associated with these gaugings is completely determined by a real two-index symmetric tensor $\theta$ in the \textbf{36}$^\prime$ of SL(8,${\mathbb R}$) and another real two-index symmetric tensor $\xi$ in the \textbf{36}. As shown in \cite{DallAgata:2011aa}, it is not restrictive to take both $\theta$ and $\xi$ diagonal. In terms of $\theta$ and $\xi$,  the quadratic constraint ensuring the consistency of the gauging reads
\begin{equation}
\theta \ \xi = {\rm (constant)} \ {\mathbb 1}_8 \, . 
\end{equation}
It is actually worth pointing out that, by suitably parameterizing $\theta$ and $\xi$, we can define a larger class of gaugings, which include those with a Minkowski vacuum we are interested in, but also many more vacua with both negative and positive cosmological constants.

\subsubsection{SO(6,2) gauging}

We consider here the class of gaugings described by Eq.~(4.27) of Ref.~\cite{DallAgata:2011aa}, whose gauge group is SO(6,2) $\subset$ SL(8,${\mathbb R}$) $\subset$ E$_{7,7}$. With respect to \cite{DallAgata:2011aa}, we choose a different parameterization of the diagonal real matrices $\theta$ and $\xi$, with $\det \theta = \det \xi = 1$, which makes clear that we are dealing with three independent parameters only:
\begin{equation}
\theta = \xi^{-1} = {\rm diag} \ 
\left( 
- \frac{1}{xyz} \, , 
- \frac{1}{xyz} \, , 
x \, , 
x \, , 
y \, , 
y \, , 
z \, , 
z 
\right) \, .
\end{equation}
The three parameters $(x,y,z)$ describing the gauging correspond to flat directions of the classical potential, along which supersymmetry is completely broken: it is not restrictive to take them real and positive. As discussed in \cite{DallAgata:2011aa}, we could add a fourth parameter corresponding to a common rescaling of $\theta$ and 
$\xi$, which can be reabsorbed in the definition of the gauge coupling constant. As it is clear from the scalar mass spectrum computed in \cite{DallAgata:2011aa}, the parameters $(x,y,z)$ are associated with 3 of the moduli fields preserving the Minkowski vacuum. However, there can be other moduli, which cannot be identified with parameters in the definition of $\theta$ and $\xi$: they are of course not included in the following considerations.

For generic values of the parameters, the unbroken gauge group is U(1)$^4$. For special values of the parameters, the unbroken gauge group can be a larger subgroup of SO(6) $\times$ SO(2) $\subset$ SO(6,2), with a lower number of explicit U(1) factors. The spectrum is given by Eqs.~(\ref{U1spectrum}) and (\ref{U1metric}), with the charge assignments of Eq.~(\ref{so62charges}) and:
\begin{eqnarray}
\mu_1^2  & = & 
\frac{(x - y) (x - z) (1 + x^2 y z)}{8 x^2 y z} \, , 
\label{so62m1sq} \\
\mu_2^2 & = & 
\frac{(y - x) (y - z) (1 + x y^2 z)}{8 x y^2 z} \, , 
\label{so62m2sq} \\
\mu_3^2  & = & 
\frac{(x - z) (y - z) (1 + x y z^2)}{8 x y z^2} \, , 
\label{so62m3sq} \\
\mu_4^2  & = & 
\frac{(1 + x^2 y z) (1 + x y^2 z) (1 + x y z^2)}{8 x^2 y^2 z^2} \, .
\label{so62m4sq} 
\end{eqnarray}
The eight massive gravitinos are always degenerate in mass. For generic values of the parameters, also the physical spin-1/2 fermions are all massive, but there are four massless vectors, associated with the unbroken U(1)$^4$. However, there are also some tachyons in the classical spectrum: the scalar states have squared masses $16 \, \mu_1^2$  for $[ijkl] = [1357], \, [2468]$, $16 \, \mu_2^2$ for  $[ijkl] = [1368], \, [2457]$  and $16 \, \mu_3^2$ for  $[ijkl] = [1458], \, [2367]$. From the explicit expressions of $\mu_1^2$, $\mu_2^2$ and $\mu_3^2$ in Eqs.~(\ref{so62m1sq}), (\ref{so62m2sq}) and (\ref{so62m3sq}), it is obvious that they cannot be simultaneously positive. To avoid tachyons, it is not restrictive to choose $z=y$. This leaves us with only two independent parameters, $x$ and $y$, and implies:
\begin{equation}
\mu_1^2  = \frac{(x - y)^2 (1 + x^2 y^2)}{8 x^2 y^2} \, , 
\quad
\mu_2^2  = \mu_3^2  =  0 \, ,
\quad
\mu_4^2  = \frac{(1 + x^2 y^2) (1 + x y^3)^2}{8 x^2 y^4} \, .
\label{mIsqnt}
\end{equation}
Notice that the unbroken gauge group on the vacuum is enhanced to SO(4) $\times$ U(1)$^2$ and, simultaneously, the mass parameters associated with the two commuting U(1) in SO(4) are set to zero. This implies that the most general tachyon-free spectrum for this class of models has: 8 massless vectors and 20 massive vectors, the latter with 3 different mass eigenvalues and degeneracies (4,8,8); in the scalar sector, there are now 20 Goldstone bosons, 18 physical massless scalars and 32 massive scalars, the latter with 5 different mass eigenvalues and degeneracies (2,2,8,8,12). Notice also that now we can identify a combination of the residual parameters $(x,y)$, which can be used for a universal rescaling of all masses, and therefore can be reabsorbed in the definition of the gauge coupling constant.

For $x=y=z$, the unbroken gauge group on the vacuum is SO(6) $\times$ U(1) and $\mu_1^2=\mu_2^2=\mu_3^2=0$: even if the two models have different interactions, we recover the same spectrum of the CSS gauging in the special case where $e_1=e_2=e_3=e_4$. 

We verified numerically that, for all parameter choices that lead to non-tachyonic classical vacua, the one-loop effective potential $V_1$ is negative definite.

\subsubsection{SO(2,2) $\times$ SO(4) $\ltimes T^{16}$}

We move now to the class of gaugings described by Eq.~(4.28) of Ref.~\cite{DallAgata:2011aa}. Also in this case we choose a different parameterization:
\begin{equation}
\theta = {\rm diag} \ 
\left( 
1 \, , 
1 \, , 
x y \, , 
x y \, , 
0 \, , 
0 \, , 
0 \, , 
0 
\right) \, , 
\qquad
\xi  = {\rm diag} \ 
\left( 
0 \, , 
0 \, , 
0 \, , 
0 \, , 
x \, , 
x \, , 
- y  \, , 
- y 
\right) \, ,
\end{equation}
and compute the mass spectrum at the origin of moduli space. The two real and positive parameters $(x,y)$ describing the gauging correspond to flat directions of the classical potential, along which supersymmetry is completely broken. As in the previous example, we could add a third parameter corresponding to a common rescaling of $\theta$ and $\xi$: this time it would be associated with another modulus that universally rescales all masses and can be reabsorbed in the gauge coupling constants. The spectrum is again determined by the charge assignments of Eq.~(\ref{so62charges}), but this time with:
\begin{equation}
\mu_1^2 =  
\frac{y  (x + y)}{8} \, , 
\quad
\mu_2^2 =  
\frac{x  (x + y)}{8} \, , 
\quad
\mu_3^2 =  
\frac{ x  y  (-1 + x y)}{8} \, , 
\quad
\mu_4^2 =  
\frac{1 - x y}{8} \, .
\end{equation}
Notice that, for generic values of the parameters $(x,y)$, the spectrum contains tachyons, since $\mu_3^2$ has the opposite sign of $\mu_4^2$.  To remove the tachyons, one possibility is to choose $y=1/x$, which leads to the gauge group SO(2,2) $\times$ SO(4) $\ltimes T^{16}$, broken to U(1)$^2$ $\times$ SO(4) on the vacuum, and to
%
\begin{equation}
\mu_1^2 = 
\frac{1 + x^2}{8 \, x^2} \, , 
\quad
\mu_2^2 =  
\frac{1 + x^2}{8} \, , 
\quad
\mu_3^2 =  \mu_4^2 =  0 \, .
\end{equation}
The other possible choice is the limit $y \rightarrow 0$, which indeed leads to a different gauging, with gauge group  U(1)$^2$ $\ltimes$ T$^{20}$ (in this case there are 6 additional spectator Abelian vector fields), broken to U(1)$^2$ on the vacuum, and to 
\begin{equation}
\mu_1^2 =  0 \, , 
\quad
\mu_2^2 =  
\frac{x^2}{8} \, , 
\quad
\mu_3^2 =  0 \, , 
\quad
\mu_4^2 =  \frac{1}{8} \, .
\end{equation}

Again, we checked numerically that, for all parameter choices that lead to non-tachyonic classical vacua, the one-loop effective potential $V_1$ is negative definite.



\section{Open questions and outlook} 
\label{subsec:Outlook}

The main results of this paper, dealing with $N=8$ supergravity in four dimensions, have already been summarized in the introductory section. We list here a number of open questions, whose answer goes beyond the scope of the present paper and our current understanding.

In Section~3 we proved the finiteness of the one-loop effective potential $V_1$ and the vanishing of ${\rm Str}\, {\cal M}^2$ and ${\rm Str}\, {\cal M}^4$ along the flat directions of the classical potential. We obtained these results in full generality: they apply not only to the known $N=8$ gaugings leading to Minkowski vacua with fully broken supersymmetry, but also to the additional gaugings that are likely to be found in the near future, generalizing the results and the techniques of  \cite{DallAgata:2011aa}. On the other hand, in Section~4 we were able to prove the vanishing of ${\rm Str}\, {\cal M}^6$ and the positivity of ${\rm Str}\, {\cal M}^8$, and to find evidence for the negativity of the one-loop effective potential $V_1$, only under two assumptions: 1) there is at least one unbroken U(1) factor in the gauge group; 2) the spectrum is determined by the charges of the supersymmetry generators with respect to the unbroken U(1) factors in the gauge group. Since all the known gaugings that fully break supersymmetry on a flat background fulfill these two assumptions, it would be interesting to see whether there are counterexamples, or whether the two assumptions are not really assumptions and can be proven on general grounds. Producing new classes of gaugings, along the lines of \cite{DallAgata:2011aa}, can be used to test the generality of the two assumptions and to shed light on the final answer. 

Even within the class of gaugings considered in Section~4, or their known explicit realizations of subsections 4.1 and 4.2, it would be important to understand whether there is some explanation of the intriguing fact that the one-loop corrections to the classically flat directions of the potential are always negative. Since there is evidence that all the $N=8$ gaugings can be associated with flux compactifications of eleven-dimensional supergravity, once geometrical and non-geometrical fluxes are introduced, it would be interesting to compute the one-loop effective potential in the compactified higher-dimensional theory (at least in the controllable cases with geometrical fluxes), and compare it with $V_1$ as computed in the four-dimensional reduced theory. Some of these gaugings can also be generated from suitable compactifications of Type-II superstrings, and also in that case it was found \cite{Rohm:1983aq} (see also \cite{Angelantonj:2006ut} and references therein) that for large enough volume there are no tachyons and $V_1 <0$.

More modestly, it would be interesting to turn the partial analytical results of  Section~4 on the correlation between ${\rm Str}\, {\cal M}^6 = 0$, ${\rm Str}\, {\cal M}^8 > 0$ and $V_1 < 0$ into a general proof, applicable to arbitrary mass spectra.

Our findings, and their possible generalizations, may provide some guidance in addressing some issues that are relevant in a wider context. While $N=8$ supergravity vacua with negative cosmological constant have been widely discussed, there are only very few examples of gaugings admitting critical points with a positive or vanishing classical  vacuum energy (for a recent investigation of stability without supersymmetry in $N=8$ supergravity, see \cite{Borghese:2011en}). Actually, despite extensive searches, there are no examples of locally stable classical de Sitter vacua, not only in $N=8$, but also in all $N > 2$ four-dimensional supergravities. One hope is that de Sitter vacua with small positive cosmological constant can be generated by quantum corrections to classical Minkowski vacua. Our study shows that, so far, in $N=8$ supergravity there is no example of one-loop stable supersymmetry-breaking vacuum with positive or vanishing vacuum energy.

Our results could also be helpful in providing a broader framework for interpreting the findings of  \cite{Grigoryan:2012xv}: in the context of $N=8$ supergravity, this paper applied some general sum rules to the CSS gauging of subsection 4.1, which admits a single non-trivial unbroken U(1) factor in the gauge group, and used them to draw conclusions on the ultraviolet incompleteness of the theory. It would be interesting to extend the interpretation of these results also to the models of subsection 4.2, where there is more than a single unbroken U(1) factor in the gauge group. 

We hope to come back to some of these points in some future publication.


\newpage

\bigskip
\section*{Acknowledgments}

\noindent We would like to thank F.~Catino, L.~J.~Dixon, S.~Ferrara, F.~Feruglio, G.~Inverso, M.~Trigiante and G.~Villadoro for helpful discussions and  C.~Angelantonj, H.~Samtleben for useful correspondence. This work is supported in part by the ERC Advanced Grants no.~226455, \textit{``Supersymmetry, Quantum Gravity and Gauge Fields''} (\textit{SUPERFIELDS}), and no.~267985, \textit{``Electroweak Symmetry Breaking, Flavour and Dark Matter: One Solution for Three Mysteries''}  (\textit{DaMeSyFla}), by the European Programme UNILHC (contract PITN-GA-2009-237920), by the Padova University Project CPDA105015/10, by the MIUR-PRIN contract 2009-KHZKRX and by the MIUR-FIRB grant RBFR10QS5J \emph{``String Theory and Fundamental Interactions''}.


\bigskip

\appendix

\section{Quadratic and quartic identities}

The quadratic constraints on the embedding tensor are necessary conditions for the consistency of the gauging procedure.
Since the $T$-tensor and the fermion shift tensors $A_1$ and $A_2$ are constructed by dressing the structure constants with the coset representatives, they are also subject to quadratic constraints derived by suitable multiplications of (\ref{contractedquadraticconstraint}) with ${\cal V}$.
The resulting equations were collected in equation (3.30) of \cite{de Wit:2007mt}, which we report here as a starting point for our discussion on the quadratic and quartic identities we used in the main text:
\begin{eqnarray}
  \label{eq:quadratic2}
Y^{k m}_{l n} &=&   A^k{}_{lij} \,A_{n}{}^{mij} - A_{l}{}^{kij} \, A^m{}_{nij}
  -4\, A^{(k}{}_{lni}A^{m)i}-4\, A_{(n}{}^{mki}A_{l)i} \nonumber  \\
  && {}-2\delta_{l}^{m}\,A_{ni}A^{ki}+2\, \delta_{n}^{k}\, A_{li}A^{mi}
  =0\,, \label{Yident}\\[3mm]
S_{[mnpq]j}^i &=&  A^i{}_{jk[m} \,A^k{}_{npq]}
  +A_{jk}\delta^{i}_{[m}\, A^{k}{}_{npq]}
  -A_{j[m}A^{i}{}_{npq]} \nonumber \\[3mm]
 && {}+\frac1{24}\varepsilon_{mnpqrstu}\,
  (A_{j}{}^{ikr}\, A_{k}{}^{stu}
  +A^{ik}\delta_{j}^{r}A_{k}{}^{stu}-A^{ir}A_{j}{}^{stu})= 0  \,,
   \label{Sident} \\[3mm]
 Z_i{}^j &=&9\, A^m{}_{ikl} \, A_{m}{}^{jkl} - A^j{}_{klm}\,A_{i}{}^{klm} -
\delta_i{}^j\, 
A^n{}_{klm} \, A_{n}{}^{klm}= 0 \,,\label{36ident} \\[3mm]
X_{ijk}^{mnp} &=& A^r{}_{ijk} \,A_{r}{}^{mnp} - 9\,A^{[m}{}_{r[ij}\,
A_{k]}{}^{np]r}
- 9\, \delta_{[i}{}^{[m} \, A^n{}_{rsj}\,A_{k]}{}^{p]rs} \nonumber \\[3mm]
&&- 9\, \delta_{[i\,j}{}^{[mn}\,A^u{}_{k]rs}\, A_{u}{}^{p]rs}
+\delta_{\,i}^m{}_j^n{}_k^p \,A^u{}_{rst} \, A_{u}{}^{rst} = 0 \,.
 \label{Xident}
\end{eqnarray}
In terms of SU(8) representations, the content of these equations is given by the representations $\mathbf{945} + \overline {\mathbf{945}} + \mathbf{63}$ for $Y^{km}_{ln}$, $\mathbf{3584}+\mathbf{378}+\overline {\mathbf{378}}+\mathbf{70}$ for $S^i_{[mnpq]j}$, $\mathbf{63}$ for $Z_i{}^j$ and $\mathbf{2352}$ for $X_{ijk}^{mnp}$.

In what follows we will discuss some combinations of the above identities that we used in the main text to simplify the computation of the quadratic and quartic supertrace formula.

\subsection{Quadratic identities} 
\label{sub:Quadratic identities}

A basic combination that we used repeatedly is the so-called supersymmetric Ward identity, which relates the matrix form of the scalar potential to the square of the shifts of the supersymmetry rules of the fermions.
This is a combination of the terms of (\ref{Yident}) and (\ref{36ident}) in the $\mathbf{63}$:
\begin{equation}
	\frac98 \,Z_i{}^j -\frac18\, Y^{k j}_{i k} = 0.
\end{equation}
Explicitly, the supersymmetric Ward identity is
\begin{equation}
	A^a{}_{jkl} A_b{}^{jkl} - 18 \, A^{ak}A_{bk}  = - \frac18 \, \delta^a_b \left(18 \, A_{ij}A^{ij} - A^i{}_{jkl}A_i{}^{jkl}\right).
	\label{piidentity}
\end{equation}
In our paper we are actually interested in Minkowski critical points and therefore we can simplify the expression of (\ref{piidentity}) by setting the terms proportional to the cosmological constant to zero:
\begin{equation}
	A^a{}_{jkl} A_b{}^{jkl} =  18 \, A^{ak}A_{bk}.
	\label{piidentity2}
\end{equation}
From the same quadratic identities, we can obtain another independent relation in the $\mathbf{63}$ representation.
Assuming a vanishing cosmological constant, the combination $\frac18\, \left(Z_i{}^j - Y^{k j}_{i k}\right)$ gives the identity
\begin{equation}
	 A^m{}_{ikl} \, A_{2m}{}^{jkl} = 2 \delta_i^j \, A_{pq} A^{pq} + 2 A_{i p} A^{p j}.
	\label{piidentity3}
\end{equation}

Additional quadratic identities can be obtained by projecting the previous ones and using the information that we all the quantities are computed at a critical point with vanishing cosmological constant.
In detail, projecting (\ref{Sident}) onto the \textbf{378+70} representations of SU(8) gives
\begin{equation}
	\label{378id}
	\epsilon_{mnpabcde} A_j{}^{kab} A_k{}^{cde} + \epsilon_{mnpjabcd} A^{ak} A_{k}{}^{bcd} =  
	-24 \,A_{jk} A^k{}_{mnp} - 18 A^t_{k[jm}A^k{}_{np]t}
\end{equation}
and its complete antisymmetrization gives the constraint in the \textbf{70}:
\begin{equation}
	\frac34 \, \epsilon_{mnpjabcd} A_e{}^{kab} A_k{}^{cde} - \epsilon_{mnpjabcd} A_{k}{}^{abc} A^{dk} =  
	-24 \,A^k{}_{[mnp}A_{j]k} + 18 A^t_{k[mn}A^k{}_{pj]t}.
\end{equation}
It is extremely interesting to note that this condition matches the structure of the critical point condition (\ref{vacuum}) but with different signs.
This means that \emph{on the vacuum} we must fulfill separately
\begin{eqnarray}
	\label{70ident}
	\frac34 \, \epsilon_{mnpjabcd} A_e{}^{kab} A_k{}^{cde} =
	-24 \,A^k{}_{[mnp}A_{j]k} , \\[2mm]
	\epsilon_{mnpjabcd} A_{k}{}^{abc} A^{dk} =  
	- 18 A^t_{k[mn}A^k{}_{pj]t}.
\end{eqnarray}
Plugging the second condition into (\ref{378id}) we eventually get that
\begin{equation}
	\epsilon_{mnpabcde} A_j{}^{kab} A_k{}^{cde} =  
	-24 \,A_{jk} A^k{}_{mnp}.
	\label{finalidentity}
\end{equation}


\subsection{Quartic identities} 
\label{sub:Quartic identities}

For the computation of the quartic supertrace, we also need identities that are quartic in the $A$-tensors and SU(8) singlets. These identities follow from contracting the quadratic identities above with other quadratic combinations of $A_1$ and $A_2$ so that the final result has all the indices contracted. We give here only the contractions we used in the main text, stressing once more that all these identities are computed also using the critical point condition (\ref{vacuum}) and setting $\Lambda = 0$. A more complete set of identities can be found in \cite{Borghese:2011en}, where a similar technique was used to investigate the classical stability of non-supersymmetric vacua with a non-vanishing cosmological constant.

The first set of identities is obtained by simple contractions of (\ref{Xident}) and (\ref{Yident}):
\begin{equation}
	I_1 = \frac19\,  A_{w}{}^{i j k} A^{w}{}_{m n p}\, X_{i j k}^{m n p},
\end{equation}
\begin{equation}
	I_2 = A^c{}_{a i j} A_{b}{}^{d i j}\, Y^{a b}_{c d},
\end{equation}
\begin{equation}
	I_3 = A_{a u} A_{b}{}^{c d u}\, Y^{a b}_{c d},
\end{equation}
\begin{equation}
	I_4 = A^{c u} A^{d}{}_{a b u}\, Y^{a b}_{c d},
\end{equation}
and their combination
\begin{equation}
	I_5 = I_2 - 2 I_3 + 2 I_4.
\end{equation}

A different set comes from identifying the square of the left hand side of (\ref{70ident}) or of (\ref{finalidentity}) with the square of the right hand side of the same equations.
From (\ref{70ident}) we get
\begin{equation}
	\begin{array}{rcl}
	I_6 &=&  {A}_{a}\,^{b c d} {A}_{b}\,^{a e f} {A}^{g}\,_{c d i} {A}^{i}\,_{e f g} - 2\, {A}_{a}\,^{b c d} {A}_{b}\,^{a e f} {A}^{g}\,_{c e i} {A}^{i}\,_{d f g} \\[3mm]
	 &&- 24\, {A}_{a b} {A}_{c d} {A}^{a c} {A}^{b d} + 4\, {A}_{a b} {A}^{c d} {A}_{c}\,^{a e f} {A}^{b}\,_{d e f} = 0,
 	\end{array}
\end{equation}
while from (\ref{finalidentity}) we get:
\begin{equation}
\begin{array}{rcl}
	I_7 &=& - 36\, {A}_{a b} {A}_{c d} {A}^{a c} {A}^{b d} + 12\, {A}_{a b} {A}_{c d} {A}^{a b} {A}^{c d} - 2\, {A}_{a}\,^{b c d} {A}_{b}\,^{e f g} {A}^{a}\,_{c e i} {A}^{i}\,_{d f g} \\[3mm]
	&&+  {A}_{a}\,^{b c d} {A}_{b}\,^{e f g} {A}^{a}\,_{e f i} {A}^{i}\,_{c d g} = 0.
	\end{array}
\end{equation}


\newpage

%

\begin{thebibliography}{10}


\bibitem{Bern:2011qn}   Z.~Bern, J.~J.~Carrasco, L.~J.~Dixon, H.~Johansson and R.~Roiban,  ``Amplitudes and Ultraviolet Behavior of N = 8 Supergravity,'' Fortsch.\ Phys.\  {\bf 59} (2011) 561 [arXiv:1103.1848 [hep-th]].

\bibitem{Coleman:1973jx}   S.~R.~Coleman and E.~J.~Weinberg,   ``Radiative Corrections as the Origin of Spontaneous Symmetry Breaking,''
  Phys.\ Rev.\ D {\bf 7} (1973) 1888.

\bibitem{Weinberg:1973ua}   S.~Weinberg,   ``Perturbative Calculations of Symmetry Breaking,''
  Phys.\ Rev.\ D {\bf 7} (1973) 2887.

\bibitem{Zumino:1974bg}  B.~Zumino,   ``Supersymmetry and the Vacuum,''
  Nucl.\ Phys.\ B {\bf 89} (1975) 535.

\bibitem{Ferrara:1979wa}   S.~Ferrara, L.~Girardello and F.~Palumbo,  ``A General Mass Formula In Broken Supersymmetry,''
  Phys.\ Rev.\  D {\bf 20} (1979) 403.

\bibitem{Girardello:1981wz}  L.~Girardello and M.~T.~Grisaru,  ``Soft Breaking of Supersymmetry,''
  Nucl.\ Phys.\ B {\bf 194} (1982) 65.

\bibitem{Grisaru:1982sr}  M.~T.~Grisaru, M.~Rocek and A.~Karlhede,  ``The Superhiggs Effect In Superspace,''
  Phys.\ Lett.\  B {\bf 120} (1983) 110.

\bibitem{Ferrara:1987jq}  S.~Ferrara, C.~Kounnas, M.~Porrati and F.~Zwirner,  ``Effective Superhiggs And Str ${\cal M}^2$ From Four-Dimensional Strings,''
  Phys.\ Lett.\  B {\bf 194} (1987) 366.

\bibitem{Ferrara:1994kg}  S.~Ferrara, C.~Kounnas and F.~Zwirner,  ``Mass formulae and natural hierarchy in string effective supergravities,''
  Nucl.\ Phys.\  B {\bf 429} (1994) 589
  [Erratum-ibid.\  B {\bf 433} (1995) 255]
  [arXiv:hep-th/9405188].

\bibitem{Cremmer:1979uq}  E.~Cremmer, J.~Scherk and J.~H.~Schwarz,  ``Spontaneously Broken N=8 Supergravity,''
  Phys.\ Lett.\ B {\bf 84} (1979) 83.

\bibitem{Ferrara:1979fu}  S.~Ferrara and B.~Zumino,  ``The Mass Matrix Of N=8 Supergravity,''
  Phys.\ Lett.\ B {\bf 86} (1979) 279.

\bibitem{DallAgata:2011aa}  G.~Dall'Agata and G.~Inverso,  ``On the Vacua of N = 8 Gauged Supergravity in 4 Dimensions,''
  Nucl.\ Phys.\ B {\bf 859} (2012) 70  [arXiv:1112.3345 [hep-th]].

\bibitem{invthe}
G.~Inverso, ``De Sitter vacua in N=8 supergravity'', Master Thesis, a.y. 2009/2010, Padova University, 14 July 2010.

\bibitem{Dibitetto:2011gm}
  G.~Dibitetto, A.~Guarino and D.~Roest,
  ``Charting the landscape of N=4 flux compactifications,''
  JHEP {\bf 1103} (2011) 137
  [arXiv:1102.0239 [hep-th]].

\bibitem{deWit:2002vt}  B.~de Wit, H.~Samtleben and M.~Trigiante,  ``On Lagrangians and gaugings of maximal supergravities,''
  Nucl.\ Phys.\ B {\bf 655} (2003) 93
  [hep-th/0212239].

\bibitem{de Wit:2007mt}  B.~de Wit, H.~Samtleben and M.~Trigiante,  ``The Maximal D=4 supergravities,''
  JHEP {\bf 0706} (2007) 049
  [arXiv:0705.2101 [hep-th]].

\bibitem{deWit:1977fk}
  B.~de Wit and D.~Z.~Freedman,
  ``On SO(8) Extended Supergravity,''
  Nucl.\ Phys.\ B {\bf 130} (1977) 105.

\bibitem{Cremmer:1978ds}  E.~Cremmer and B.~Julia,  ``The N=8 Supergravity Theory. 1. The Lagrangian,''
  Phys.\ Lett.\ B {\bf 80} (1978) 48.

\bibitem{Cremmer:1979up}  E.~Cremmer and B.~Julia,  ``The SO(8) Supergravity,''
  Nucl.\ Phys.\ B {\bf 159} (1979) 141.

\bibitem{de Wit:1982ig} 
  B.~de Wit and H.~Nicolai,
  ``N=8 Supergravity,''
  Nucl.\ Phys.\ B {\bf 208}, 323 (1982).

\bibitem{Nicolai:2000sc}  H.~Nicolai and H.~Samtleben,  ``Maximal gauged supergravity in three-dimensions,''
  Phys.\ Rev.\ Lett.\  {\bf 86} (2001) 1686
  [hep-th/0010076].

\bibitem{Gaillard:1981rj}  M.~K.~Gaillard and B.~Zumino,  ``Duality Rotations for Interacting Fields,''
  Nucl.\ Phys.\ B {\bf 193} (1981) 221.
  
\bibitem{Dall'Agata:2005mj} G.~Dall'Agata, R.~D'Auria and S.~Ferrara, ``Compactifications on twisted tori with fluxes and free differential algebras,''
  Phys.\ Lett.\ B {\bf 619} (2005) 149
  [hep-th/0503122].

\bibitem{deWit:2005ub}  B.~de Wit, H.~Samtleben and M.~Trigiante,  ``Magnetic charges in local field theory,''
  JHEP {\bf 0509} (2005) 016
  [hep-th/0507289].

\bibitem{Dall'Agata:2007sr}  G.~Dall'Agata, N.~Prezas, H.~Samtleben and M.~Trigiante,  ``Gauged Supergravities from Twisted Doubled Tori and Non-Geometric String Backgrounds,''
  Nucl.\ Phys.\ B {\bf 799} (2008) 80
  [arXiv:0712.1026 [hep-th]].

\bibitem{de Wit:1981eq}  B.~de Wit and H.~Nicolai,  ``N=8 Supergravity with Local SO(8) $\times$ SU(8) Invariance,''
  Phys.\ Lett.\ B {\bf 108} (1982) 285.

\bibitem{deWit:1983gs}  B.~de Wit, H.~Nicolai,  ``The Parallelizing S(7) Torsion In Gauged N=8 Supergravity,''
  Nucl.\ Phys.\  {\bf B231 } (1984)  506.

\bibitem{LeDiffon:2011wt}  A.~Le Diffon, H.~Samtleben and M.~Trigiante,  ``N=8 Supergravity with Local Scaling Symmetry,''
  JHEP {\bf 1104} (2011) 079
  [arXiv:1103.2785 [hep-th]].

\bibitem{Peeters:2006kp}  K.~Peeters,  ``A Field-theory motivated approach to symbolic computer algebra,''
  Comput.\ Phys.\ Commun.\  {\bf 176} (2007) 550
  [cs/0608005 [cs.SC]].

\bibitem{Peeters:2007wn}  K.~Peeters,  ``Introducing Cadabra: A Symbolic computer algebra system for field theory problems,''
  hep-th/0701238 [HEP-TH].

\bibitem{D'Auria:2003jk}
  R.~D'Auria, S.~Ferrara, F.~Gargiulo, M.~Trigiante and S.~Vaula,
  ``N=4 supergravity Lagrangian for type IIB on T**6 / Z(2) in presence of fluxes and D3-branes,''
  JHEP {\bf 0306} (2003) 045
  [hep-th/0303049].

\bibitem{Andrianopoli:2002rm}
  L.~Andrianopoli, R.~D'Auria, S.~Ferrara and M.~A.~Lledo,
  ``Super Higgs effect in extended supergravity,''
  Nucl.\ Phys.\ B {\bf 640} (2002) 46
  [hep-th/0202116].

\bibitem{Villadoro:2004ec}
  G.~Villadoro,
  ``N = 6 gauged supergravities from generalized dimensional reduction,''
  Phys.\ Lett.\ B {\bf 602} (2004) 123
  [hep-th/0407105].

\bibitem{Scherk:1979zr}  J.~Scherk and J.~H.~Schwarz,  ``How to Get Masses from Extra Dimensions,''
  Nucl.\ Phys.\ B {\bf 153} (1979) 61.

\bibitem{D'Auria:2005er}  R.~D'Auria, S.~Ferrara, M.~Trigiante,  ``E$_{7(7)}$ symmetry and dual gauge algebra of M-theory on a twisted seven-torus,''
  Nucl.\ Phys.\  {\bf B732 } (2006)  389-400.
  [hep-th/0504108].

\bibitem{D'Auria:2005dd}   R.~D'Auria, S.~Ferrara, M.~Trigiante,  ``Curvatures and potential of M-theory in D=4 with fluxes and twist,''
  JHEP {\bf 0509 } (2005)  035.
  [hep-th/0507225].

 \bibitem{D'Auria:2005rv}   R.~D'Auria, S.~Ferrara, M.~Trigiante,  ``Supersymmetric completion of M-theory 4D-gauge algebra from twisted tori and fluxes,''
  JHEP {\bf 0601 } (2006)  081.
  [hep-th/0511158].

\bibitem{Sezgin:1981ac}   E.~Sezgin and P.~van Nieuwenhuizen,  ``Renormalizability Properties Of Spontaneously Broken N=8 Supergravity,''
  Nucl.\ Phys.\  B {\bf 195} (1982) 325.

\bibitem{Sezgin:1982pk}   E.~Sezgin and P.~van Nieuwenhuizen,  ``Ultraviolet Finiteness Of N=8 Supergravity, Spontaneously Broken By
  Dimensional Reduction,''
  Phys.\ Lett.\  B {\bf 119} (1982) 117.

\bibitem{Andrianopoli:2002vy}   L.~Andrianopoli, R.~D'Auria, S.~Ferrara and M.~A.~Lledo,  ``Gauged extended supergravity without cosmological constant: No scale structure and supersymmetry breaking,''
  Mod.\ Phys.\ Lett.\ A {\bf 18} (2003) 1001
  [hep-th/0212141].

\bibitem{CatDalInvZwi} F. Catino, G. Dall'Agata, G. Inverso and F. Zwirner, to appear.

\bibitem{Rohm:1983aq}  R.~Rohm,  ``Spontaneous Supersymmetry Breaking in Supersymmetric String Theories,''
  Nucl.\ Phys.\ B {\bf 237} (1984) 553.

\bibitem{Angelantonj:2006ut}
  C.~Angelantonj, M.~Cardella and N.~Irges,
  ``An Alternative for Moduli Stabilisation,''
  Phys.\ Lett.\ B {\bf 641} (2006) 474
  [hep-th/0608022].

\bibitem{Borghese:2011en}  A.~Borghese, R.~Linares and D.~Roest,  ``Minimal Stability in Maximal Supergravity,''
 JHEP {\bf 1207} (2012) 034 [arXiv:1112.3939 [hep-th]].

\bibitem{Grigoryan:2012xv}  H.~R.~Grigoryan and M.~Porrati,  ``New Sum Rules from Low Energy Compton Scattering on Arbitrary Spin Target,''  JHEP {\bf 1207} (2012) 048  [arXiv:1204.1064 [hep-th]].

%
\end{thebibliography}
\end{document}